\documentclass[12pt]{article}

\usepackage{graphicx}
\def\be{\begin{equation}}
\def\ee{\end{equation}}
\newcommand{\beq}{\begin{equation}}
\newcommand{\eeq}[1]{\label{#1} \end{equation}}

\begin{document}

\title{\Large \bf Bjorken model with Freeze Out}
\author{\large V.K. Magas$^1$, L.P. Csernai$^{2,3}$ \bigskip \\
{\it  $^1$~Departament d'Estructura i Constituents de la Mat\'eria,}\\
{\it  Universitat de Barcelona, Diagonal 647, 08028 Barcelona, Spain} \\  
{\it $^2$~Theoretical and Energy Physics Unit,}\\ 
{\it University of Bergen, Allegaten 55, 5007 Bergen, Norway} \\ 
{\it $^3$~MTA-KFKI, Research Inst of Particle and Nuclear Physics,}\\
{\it H-1525 Budapest 114, P.O.Box 49, Hungary}}

\maketitle

{\large

\begin{center}
{\bf Abstract}\\
\medskip
The freeze out of the expanding systems, created in relativistic heavy ion collisions, is discussed. We combine Bjorken scenario with earlier developed freeze out equations into a unified model. The important feature of the proposed model is that physical freeze out is 
completely finished in a finite time, which can be varied from $0$ (freeze out hypersurface) to 
$\infty$. The dependence of the post freeze out distribution function on this freeze out time will be studied. As an example model is completely solved and analyzed for the gas of
pions.  
\end{center}


In the ultrarelativistic heavy ion collisions at RHIC the total number of the produced particles exceeds several thousands, 
therefore one can expect that the produced system behaves as a "matter" and generates collective effects. Indeed strong collective flow patterns have been measured at RHIC, 
which suggests that  
the hydrodynamical models are well justified during the intermediate stages of the reaction: from the time when local equilibrium is reached until the freeze out (FO),
when the hydrodynamical description breaks down.

In the recent works \cite{ModifiedBTE} it has been shown that the basic assumptions of the Boltzman Transport Equation (BTE) are not satisfied during FO and therefore its description has to be based on Modified BTE,  suggested in \cite{ModifiedBTE}.
When the characteristic length scale, describing the change of the distribution function, becomes smaller than mean free path (this always happens at late stages of the FO), then the basic
assumptions of the BTE get violated, and the expression for the collision integral has
to be modified to follow the trajectories of colliding particles, what makes calculations much more complicated.  
In fact, in cascade models one
follows the trajectories of the colliding particles, therefore what is effectively solved is not BTE, but Modified BTE.
Once the necessity of the Modified BTE and at the same time the difficulty of its direct solution were realized, the simplified kinetic FO models became even more important for the understanding of the principal features of these phenomenon. 

In simulations FO is usually described in two extreme ways: either FO on a hypersurface with zero thickness, or FO described by volume emission model or hadron cascade, which  require an infinite time and space for a complete FO. 
However recently a new type of FO models has been developed in Refs. \cite{Mo05a,Mo05b}, in which freeze out is gradual and completely finished in a finite layer. Thickness of this FO layer can be varied from $0$ (freeze out hypersurface) to $\infty$, and the dependence of the post freeze out distribution function on this freeze out time will be studied, making, thus, a bridge between the two extreme FO schemes discussed above. 

The model in Refs. \cite{Mo05a,Mo05b} is a further development of simplified FO models studied in Refs. \cite{old_SL_FO,old_TL_FO}, but in all these papers the expansion of the system was neglected. Thus, there is an important question to be studied: whether the important freeze out features, pointed out in these works, are not smeared out by the expansion of the system?  

In this paper we present a simple kinetic FO model, which describes the freeze out of particles from a Bjorken expanding fireball.
Thus, this is a more physical extension of the oversimplified FO models without expansion \cite{old_SL_FO,old_TL_FO,Mo05a,Mo05b}. 
Taking the basic ingredients of the FO simulations from Refs. \cite{Mo05a,Mo05b} we can make gradual freeze out to be  completely finished in a layer, i.e. in a domain restricted by two parallel 
hypersurfaces $\tau=\tau_1$ and $\tau=\tau_1+L$, where $\tau$ is the proper time, $\tau=\sqrt{t^2-x^2}$ \cite{Bjorken_FO,Bjorken_FO_mass}. Thus, the evolution of the fireball created in relativistic heavy ion collision goes as follows.\\
{\bf Initial state, $\tau=\tau_0$:}\ \  $e_0,\ n_0$\\
{\bf Phase I, Pure Bjorken hydrodynamics, $\tau_0\le \tau\le \tau_1$}
The evolution of the energy density and baryon density is given by the following equations  \cite{Bjorken}:
\beq
d e/ \tau=-(e+P)/\tau\,, \quad d n/d \tau=-n/\tau\,,
\eeq{Bjorken}   
where $P$ is the pressure. Bjorken model describes 1D directed process - only the proper time gradients are considered, changes in the other directions are neglected. 
This system can be easily solved:
\beq
e(\tau)=e_0\left(\tau_0/\tau\right)^{1+c_o^2}\,, \quad
n(\tau)=n_0\left(\tau_0/\tau\right)\,,
\eeq{pure_bjor}
where $P=c_o^2 e$ is the equation of state (EoS) in general form. In order to have have a finite volume fireball, we need to put some boarders on the system. Here we assume that our system, described by the 
Bjorken model, is situated in the spatial domain $|\eta|\le \eta_{R}$ or what is the same $|z|\le z_R(\tau) = \tau \sinh \eta_{R}$ ($\eta=\frac{1}{2}\ln \left(  \frac{t+z}{t-z}\right)$ is pseudorapidity). As the system expends the volume of the fireball increases as
$
V(\tau)=2 A_{xy} \sinh \eta_{R}\tau \,,
$
where $A_{xy}$ is the transverse area of the system.\\
{\bf Phase II, Bjorken expansion and gradual FO,   $\tau_1\le \tau\le \tau_1+L$}\\
In order to describe freeze out we introduce two components of the distribution function, $f$: the interacting, $f^i$, and the free (frozen out), $f^f$  ones, ($f=f^i+f^f$). Now we can generate the interacting and free parts of baryon flow and energy-momentum tensor based on the corresponding components of the  distribution function, which are parameterized in terms of interacting and free  energy and baryon densities correspondingly. 

According to Refs. \cite{Bjorken_FO,Bjorken_FO_mass} the evolution of such a system is govern by the following equations for interacting and free components:
\beq
\frac{d e^i}{d \tau}=-\frac{e^i+P^i}{\tau}-\frac{e^i}{\tau_{FO}}\frac{L}{L+\tau_1-\tau}\,, 
\quad 
\frac{d n^i}{d \tau}=-\frac{n^i}{\tau}-\frac{n^i}{\tau_{FO}}\frac{L}{L+\tau_1-\tau}\,, 
\eeq{int}   
\beq
\frac{d e^f}{d \tau}=-\frac{e^f}{\tau}+\frac{e^i}{\tau_{FO}}\frac{L}{L+\tau_1-\tau}\,,
\quad 
\frac{d n^f}{d \tau}=-\frac{n^f}{\tau}+\frac{n^i}{\tau_{FO}}\frac{L}{L+\tau_1-\tau}\,.
\eeq{free}
In all these equations on the r.h.s. there are two terms: the first one is due to expansion (same as in Bjroken eqs. (\ref{Bjorken})) and the second one is due to freeze out, see Refs. \cite{Bjorken_FO,Bjorken_FO_mass} for more details.

Solving eqs. (\ref{int}) we obtain: 
\beq
e^i(\tau)=e_0\left(\frac{\tau_0}{\tau}\right)^{1+c_o^2}\left(\frac{L+\tau_1-\tau}{L}\right)^{L/\tau_{FO}}\,, 
\quad n^i(\tau)=n_0\left(\frac{\tau_0}{\tau}\right)\left(\frac{L+\tau_1-\tau}{L}\right)^{L/\tau_{FO}}\,.
\eeq{bjor_FO_2}
The difference with respect to the pure Bjorken solution (\ref{pure_bjor}) is in the last multiplier, and we see that the interacting component completely disappears when $\tau$ reaches the outer edge of FO layer $\tau= L+\tau_1$.

With these last equations we have completely determined evolution of the interacting component \cite{Bjorken_FO,Bjorken_FO_mass}. Knowing $e^i(\tau)$ and EoS we can find temperature, $T_i(\tau)$. Due to symmetry of the system $u_i^\mu(\tau)=u^\mu(\tau_0)=(1,0,0,0)$. Finally, $f^i(\tau)$ is a thermal distribution with given $T_i(\tau)$, $n^i(\tau)$, $u_i^\mu(\tau)$. 

However, the most interesting for us is the free component, which is the source of the observables. Eqs. (\ref{free}) give us the evolution of the $e_f$ and $n_f$, and one can easily check that these two equations are equivalent to the following equation on the distribution function:
\beq
\frac{d f^f}{d \tau}=-\frac{f^f}{\tau}+\frac{f^i}{\tau_{FO}}\frac{L}{L+\tau_1-\tau}\,.
\eeq{ffree}
The measured post FO spectrum is given by the distribution function at the outer edge of the FO layer, i.e. by $f^f(L+\tau_1)$. 

Most of the observables will depend only on this momentum distribution. However the two particle correlations depend also on the space-time origins of the detected particles. Thus, in order to calculate two particle correlations we have to keep track on both  the momentum and the space-time coordinates of the freeze out point of particles (this can be done, for example, in the source function formalism), i.e. we need to know full evolution of free component and correspondingly can get some restrictions on it from the data.

\begin{figure*}[htb!]
\centering
\includegraphics[width=0.7\textwidth]{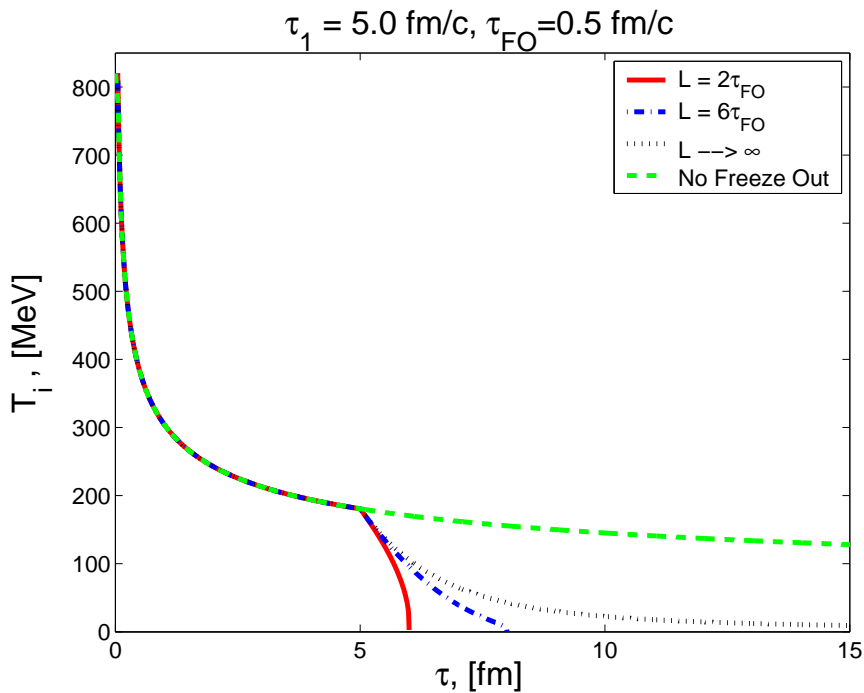}
\vspace{-0.5cm}
\caption{Evolution of temperature of the interacting matter for
different FO layers. $T_i(\tau_0=0.05\ fm)=820\ MeV$, $T_{FO}=180\ MeV$. "No Freeze Out" means that we used standard Bjorken hydrodynamics even in phase II.}
\label{fig1}
\end{figure*}


To illustrate the FO process more quantitatively we show below the results for the ideal massive pion gas with J\"uttner equilibrated distribution: 
\beq
f^i(\tau,\vec{p})=\frac{g}{(2\pi)^3}e^{-\sqrt{|\vec{p}|^2+m_\pi^2}/T_i(\tau)}\,,
\eeq{Jut}
where the degeneracy of pion is $g=3$. 
\\ \indent
Contrary to the illustrative example in \cite{Bjorken_FO} here we do not neglect the pion mass. 
During FO the temperature of the interacting component decreases to zero, so at late stages 
of the FO process this new calculation is better justified. 
We will see below that $T_i$ falls below $m_\pi$ quite soon, and so the J\"uttner distribution is a good approximation of the proper Bose pion distribution  \cite{Bjorken_FO_mass}. 
\\ \indent
For our system we have the following EoS:
\beq
e^i=\frac{3g}{2\pi^2}m^2T_i^2K_2(a)+\frac{g}{2\pi^2}m^3T_i K_1(a)\,, \quad 
P^i=\frac{g}{2\pi^2}m^2T_i^2K_2(a)\,,
\eeq{EoS}
where $K_n$ is Bessel function of the second kind, and $a=m/T_i$.

Eqs.  (\ref{int}) and (\ref{EoS}) result into the following equation for the evolution of the temperature of the interacting component:
$$
\frac{dT_i}{d \tau}= -\frac{T_i}{\tau} \frac{4 T_i^2 K_2(a) + m T_i K_1(a) }
{12 T_i^2 K_2(a) + 5 m T_i K_1(a) + m^2 K_0(a)}
$$
\beq
- \frac{T_i}{\tau_{FO}} \left(\frac{L}{L+\tau_1-\tau} \right)\frac{3 T_i^2 K_2(a) + m T_i K_1(a) }
{12 T_i^2 K_2(a) + 5 m T_i K_1(a) + m^2 K_0(a)}\,.
\eeq{T_eq}
Furthermore, we have used the following values of the parameters: $\eta_R=4.38$, $A_{xy}=\pi R_{Au}^2$, where $ R_{Au}=7.685$ fm is the $Au$ radius, $\tau_0=0.05\ fm$, $T_i(\tau_0)=820\ MeV$, $\tau_1=5\ fm$, what leads to $T_i(\tau_1)=T_{FO}=180\ MeV$, and  $\tau_{FO}=0.5\ fm$. During the pure Bjorken case the evolution of the temperature is govern by eq. (\ref{T_eq}) without the second (freeze out) term on the r.h.s. 

In Fig. \ref{fig1} we present the evolution of the temperature of the interacting matter, $T_i(\tau)$, for different values of FO time $L$. 

\begin{figure*}[htb!]
\centering
\includegraphics[width=9.8cm, height =7.8cm]{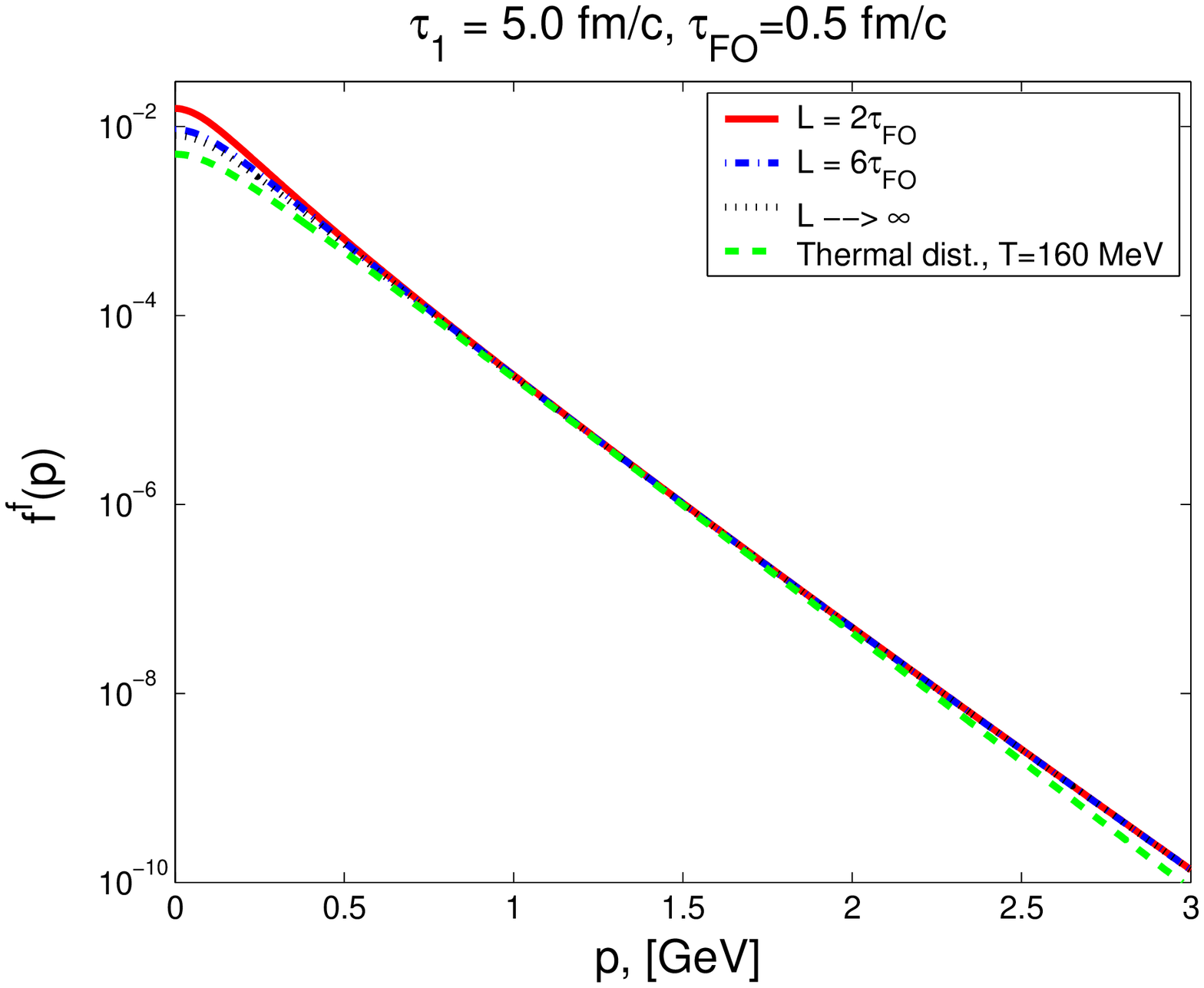} 
\includegraphics[width=6.4cm, height =7.5cm]{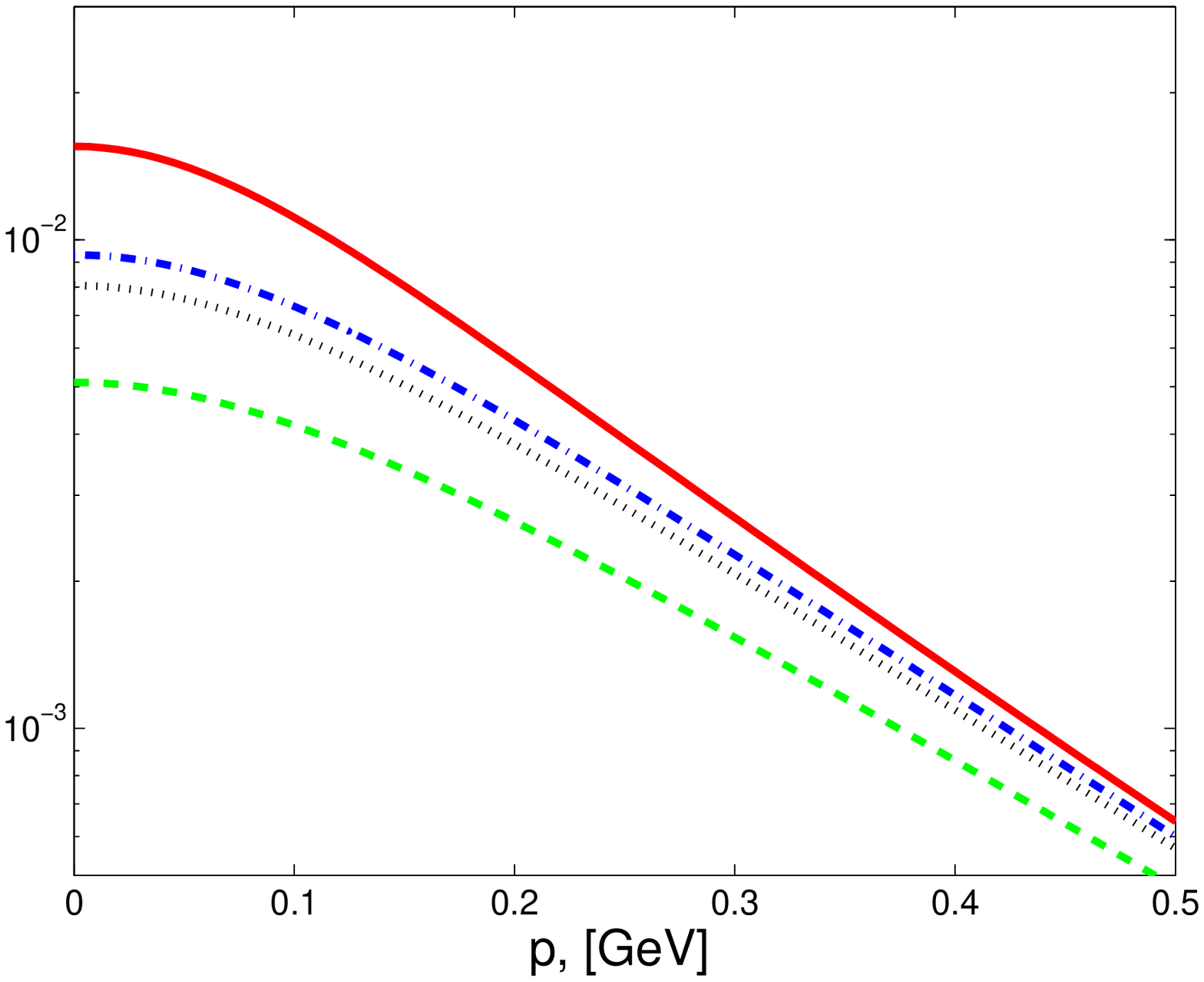}
\vspace{-0.5cm}
\caption[]{Final post FO distribution for different FO layers as a function of the momentum in the
FO direction, $p=p^x$ in our case ($p^y=p^z=0$). The initial conditions are specified in the text. Dashed curve shows thermal distribution with temperature $T=160\ MeV$.
}
\label{fig2}
\end{figure*}

As it was already shown in Ref. \cite{old_TL_FO,Mo05b}, the final post FO particle distributions, shown on Fig. \ref{fig2},
are non-equilibrated distributions, which deviate from thermal ones particularly in the low momentum region.
By introducing  and varying the thickness of the FO layer, $L$, we are strongly affecting the evolution
of the interacting component, see Fig. \ref{fig1}, but the final post FO distributions show strong universality:
for the FO layers with a thickness of several $\tau_{FO}$, the post FO distribution already looks very close to that for an infinitely long FO calculations, see Fig. \ref{fig2} left plot. Differences can be observed only for the very small momenta, as shown in  Fig. \ref{fig2} right plot.
So, the inclusion of the expansion into our consideration does not smear out this very important feature of the 
gradual FO.
\\ \indent
Please note, that if one would look on our post FO distributions only in the medium momenta regions (with $0.5\ GeV<|p|< 2.5\ GeV$), then one could fit these spectra 
reasonable well with equilibrated distribution with temperature $T=160\ MeV$, see dashed line on Fig. \ref{fig2}. However, for low and high momenta such a fit would strongly disagree.
\\ \indent
Figs. \ref{entrop} present the evolution of the total entropy, $S(\tau)$, calculated based on the full 
distribution function, $f(\vec{p})=f^i(\vec{p})+f^f(\vec{p})$:
\beq
s(\tau)=\int d^3 p f(\tau) \left[ 1 - \ln \left( \frac{(2\pi)^3}{g} f(\tau) \right)\right]\,, 
\quad
S(\tau)=s(\tau)V(\tau)\,.
\eeq{stot}
During pure Bjorken phase total entropy remains constant, as expected, but during phase II it constantly increases until FO is finished.

\begin{figure*}[htb!]
\centering
\includegraphics[width=10.0cm, height =8.00cm]{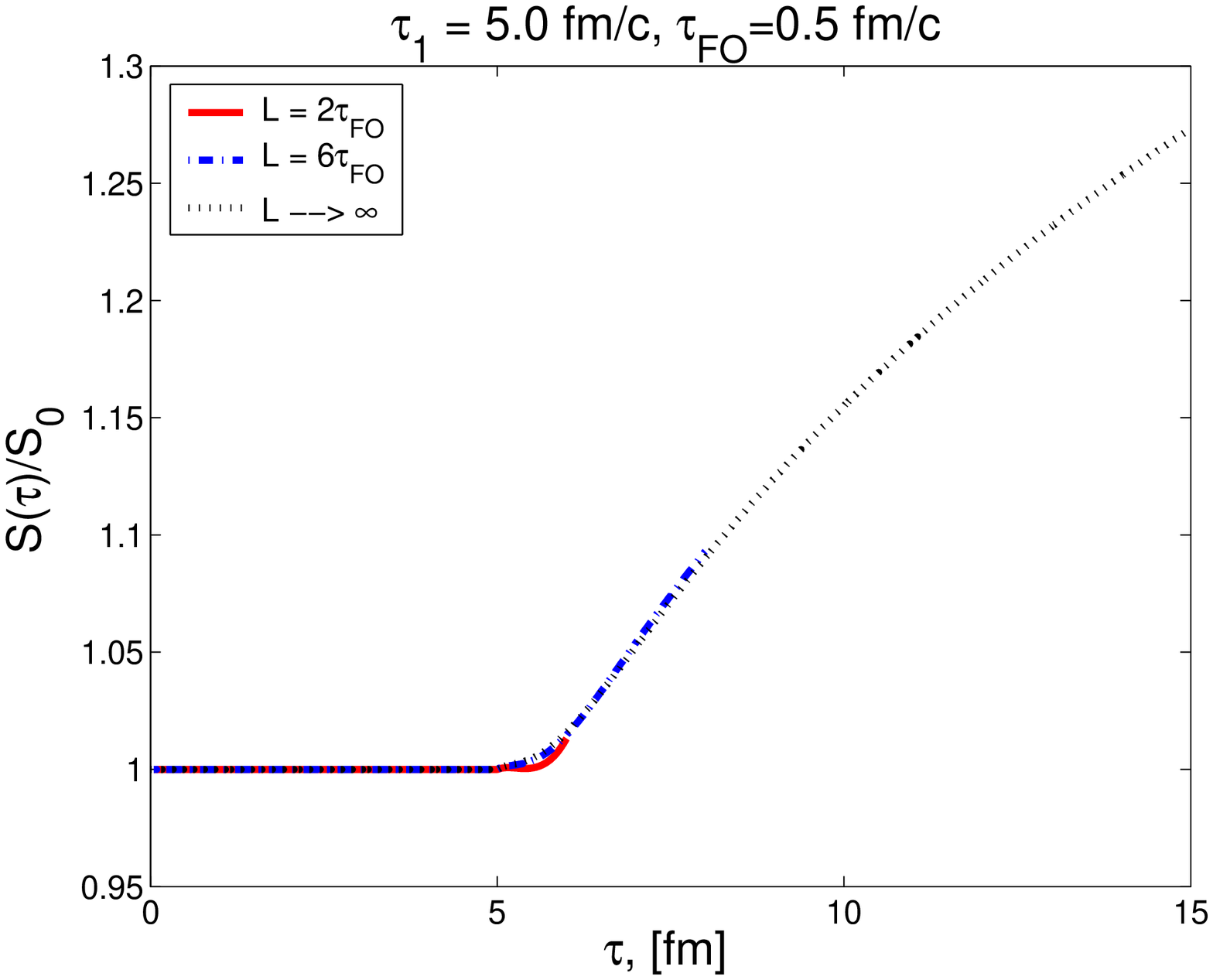} 
\includegraphics[width=6.0cm, height =7.8cm]{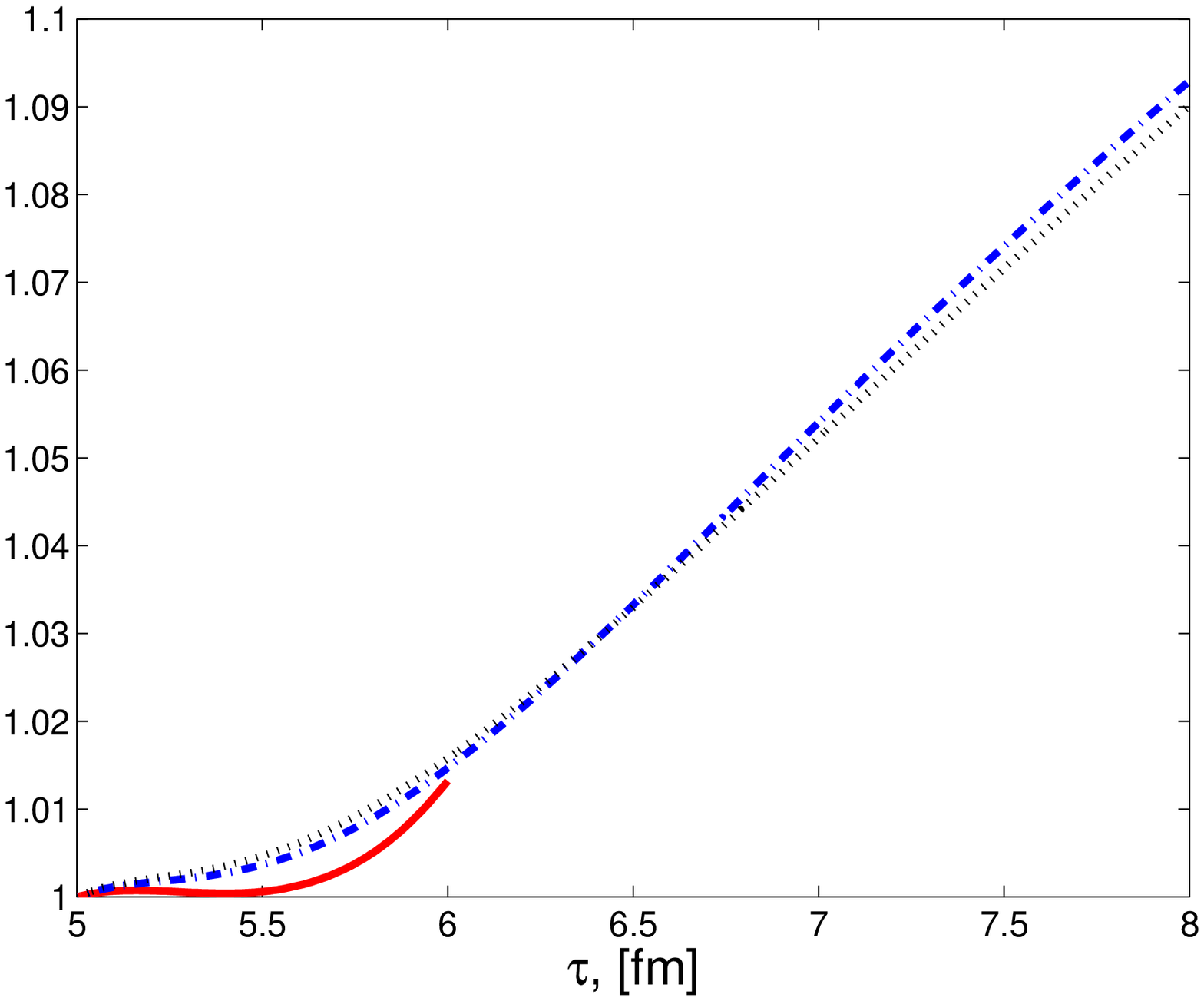}
\vspace{-0.5cm}
\caption{Evolution of the total entropy for
different FO layers.  The initial conditions are specified in the text. }
\label{entrop}
\end{figure*}

This is a very important conclusion of our model, which is stressing once again the importance to always check the non-decreasing entropy condition  \cite{cikk_2,Bjorken_FO,Bjorken_FO_mass}, since long gradual freeze out may produce substantial amount of entropy, see Fig. \ref{entrop}.

This may have important consequence for QGP search. For many qualitative estimations it was assumed that all the entropy is produced at the early stages of the reaction and that   the expansion, hadronization and FO go adiabatically, and thus number of pions can serve as a rough measure of entropy. However, if a non-negligible part of the entropy, say $10 \%$ is produced during FO, then some estimations, for example of strangeness versus entropy (pion) production \cite{Goren}
have to be reviewed. 

{\bf Aknowledgements.\ \ }Authors thank E. Molnar for fruitful discussions. This work was partially
supported by Grant No. FIS2005-03142 from MEC (Spain) and FEDER.

}

\end{document}